\documentclass[11pt,twoside]{article}
\usepackage{amssymb}
\usepackage{graphicx}
\usepackage{graphicx,type1cm,eso-pic,color}
\usepackage{amssymb}
\usepackage{amssymb,amsmath}
\usepackage{graphicx,epsfig}
\usepackage{enumerate}
\usepackage{color}
\usepackage {multicol}
\usepackage{lineno}
\numberwithin{equation}{section}
\newtheorem{theorem}{Theorem}
\newtheorem{lemma}{Lemma}

\newtheorem{example}{Example}

\makeatletter
\AddToShipoutPicture{
	\setlength{\@tempdimb}{.5\paperwidth}
	\setlength{\@tempdimc}{.5\paperheight}
	\setlength{\unitlength}{1pt}
	\put(\strip@pt\@tempdimb,\strip@pt\@tempdimc)
}
\makeatother

\begin{document}
	\setcounter{page}{1}

	\thispagestyle{empty}
	\markboth{}{}

	\pagestyle{myheadings}
	\markboth{ S.K.Chaudhary and N.Gupta }{ S.K.Chaudhary and N.Gupta }
	
	\date{}
	
	
	\noindent  
	
	\vspace{.1in}
	
	{\baselineskip 20truept
		
		\begin{center}
			{\Large {\bf Testing of symmetry based on cumulative past and residual extropy of record values}} \footnote{\noindent	{\bf *} Corresponding author E-mail: skchaudhary1994@kgpian.iitkgp.ac.in\\
				{\bf ** }  E-mail: nitin.gupta@maths.iitkgp.ac.in }\\
			
	\end{center}}

	\vspace{.1in}
	
	\begin{center}
		{\large {\bf Santosh Kumar Chaudhary* and Nitin Gupta**}}\\
		{\large {\it Department of Mathematics, Indian Institute of Technology Kharagpur, West Bengal 721302, India. }}
		\\
	\end{center}

	\vspace{.1in}
	\baselineskip 12truept
	\begin{center}
		{\bf \large Abstract}\\
	\end{center}
	In this paper, we are testing the symmetry in the distribution of data observed on a random variable. We proposed test statistics using cumulative past and residual extropy of record values based on the characterization developed by Gupta and Chaudhary (2022) \cite{guptaskc}. It is shown that the obtained estimator is consistent. Our proposed test has an advantage that we do not need to estimate the centre of symmetry. The empirical density, critical value and power of the proposed test statistics have been obtained. The test procedure has been implemented on six real-life data sets to verify its performance in identifying the symmetric nature. Simulations indicate our test performs better than the competitor tests. 

	\section{Introduction}
	Let $X_1, X_2, ... , X_N$ be a random sample of size $N$ from population $X$ with unknown probability density function(pdf) $f$. A probability distribution is said to be symmetric around $k$ if and only if there exists a finite number $k$ such that $f_X(k+x)=f_X(k-x) $ for all $x\in \mathbb{R}$  that is, $F_X(k-x)+F_X(k+x)=1$ for all $x\in \mathbb{R}$. Here, we are interested in testing whether a distribution is symmetric or skewed. That is,
	\begin{align*}
		&H_0: f_X(k+x)=f_X(k-x) \ \text{for all} \ x \in \mathbb{R}\\  \text{against} \ \ &H_1: f_X(k+x) \neq f_X(k-x)\  \text{for some} \  x\in \mathbb{R}.
	\end{align*}
	In some areas such as economy and finance, testing for symmetry of a distribution may be very important. A variety of tests for symmetry have been proposed in the literature. Interested readers can refer to McWilliams (1990) \cite{mcwill}, Gibbons and Chakraborti (1992)\cite{gibchakr92}, Tajuddin (1994)\cite{taj94}, Modarres and Gastwirth (1996)\cite{modagast}, Baklizi (2003 \cite{baklizi03}, 2007\cite{baklizi07}, 2008\cite{baklizi08}), Cheng and Balakrishnan (2004)\cite{chengbala04}, and Corzo and Babativa (2013)\cite{corzobaba} for more details on this topic. It is to be noted that all the tests are constructed for continuous distributions with a known median.
	Vasicek (1976) \cite{vasicek} provided the estimate of entropy per observation and established a goodness of fit test for normality using sample entropy. Park (1999) \cite{park} provided the sample entropy of order statistics, and presented a goodness-of-fit test for normality based on the sample entropy of order statistics. McWilliams (1990) \cite{mcwill} proposed a new test for symmetry based on runs statistics and presented an analysis of the proposed test with respect to some alternatives. Cheng and Balakrishnan (2004) \cite{chengbala04} proposed another test for symmetry based on the information on both absolute ranks and signs of the observations recorded on the random sample. Corzo and Babativa (2013) \cite{corzobaba} presented a modified version of the test proposed by Modarres and Gastwiirth (1996) \cite{modagast} for the testing symmetry of a probability distribution with a known central value.

	How can we test the distribution's symmetry if the median is unknown? Recently Xiong et al. (2021) \cite{xiongetal}  provided the property of symmetric distribution using extropy of $k$th upper and lower record value and proposed a test of symmetry based on that property. Jose and Sathar (2022) \cite{josesathar} gave a characterization result of symmetric distribution using extropy of $n$th upper $k$-record value and $n$th lower $k$-record value and proposed a test for symmetry based on that characterization.  In this paper, we present a test of symmetry using characteristics developed in Gupta and Chaudhary (2022) \cite{guptaskc} based on cumulative past and residual extropy of record values.

	Entropy was first introduced by Shannon (1948) \cite{shanon48} and it measures the average level of uncertainty related to the results	of the random experiment. The Shannon entropy for continuous random variable $X$ is defined  as:
	\begin{equation}
		\label{1eq2}
		H(X) = - {\int_{-\infty}^{\infty}{f(x) \ln \left( f(x)\right)dx}}=E\left(- \ln f(X)\right).
	\end{equation} 
	Lad et al. (2015) defined the compliment dual of the Shannon entropy called extropy. The extropy of continuous random variable $X$ is defined as:	
	\begin{equation}\label{extropy}
		J(X)=-\frac{1}{2} \int_{-\infty}^{\infty}f^2(x)dx=-\frac{1}{2}E\left(f(X)\right).
	\end{equation}
	The cumulative residual extropy  of continuous random variable $X$ is defined as :
	\begin{align}
		\xi J (X) &=- \frac{1}{2} \int_{-\infty}^{\infty} \bar{F}_{X}^2 (x) dx.
	\end{align} 
	The cumulative past extropy of $X$ is defined as:
	\begin{align}
		\bar{\xi} J (X) &=- \frac{1}{2}\int_{-\infty}^{\infty} F_{X}^2 (x) dx.
	\end{align}

	\indent The concept of $k$-records was introduced by Dziubdziela and Kopociński (1976) \cite{dzko1976}; for more details, also see  Ahsanullah (1995) and Arnold et al. (1998) \cite{arnoldbalanag98}. The pdf of the $n$th upper $k$-record value $U_{n,k}$ and the $n$th lower $k$-record value $L_{n,k}$ respectively are given by (see Arnold et al.(2008) \cite{arnoldbalanaga08} and  Ahsanullah (2004)\cite{ahsan2004}) 
	
	\begin{align*}
		&f_{U_{n(k)}}(x)=\dfrac{k^n}{(n-1)!}[-\log \overline{F}(x)]^{n-1} \overline{F}(x)^{k-1}f_X(x),\label{r1}\\
		\text{and} \ \ \ &g_{L_{n(k)}}(x)=\dfrac{k^n}{(n-1)!}[-\log F_X(x)]^{n-1} F_X(x)^{k-1}f_X(x).\\ 
	\end{align*}
	
	\noindent The cdf of $U_{n,k}$ and  $L_{n,k}$, respectively, are  
	\begin{align*}
		F_{U_{n,k}}(u)&=  1-\bar{F}^k_X (u)   \sum_{i=0}^{n-1} \frac{(-k \log\bar{F}_X (u) )^i}{i!},\\
		\text{and} \ \ \ 	F_{L_{n,k}}(u)&=  {F}^k_X (u)   \sum_{i=0}^{n-1} \frac{(-k \log F_X (u) )^i}{i!}.
	\end{align*}
	
	In light of the benefits of $k$-records over traditional records and with reference to the study of  Xiong et al.(2021) \cite{xiongetal}, Jose and Sathar (2022) \cite{josesathar}. The current study focuses on creating a test of symmetry of the distribution of a random variable based on characterization results for symmetry using cumulative and past extropy based on the $k$-records. The study has been organized as follows: Section 2 presents a characterization result (for more see, Gupta and Chaudhary (2022)\cite{guptaskc}, Ahmadi (2021) \cite{ahmadi}) for the symmetric nature of a probability distribution using cumulative and past extropy of $n$th upper and lower $k$-records. Section 3 uses the characterization result in deriving the test statistic for identifying symmetry and an estimator for the proposed test statistic is also suggested. Moreover, we note that shift transformation does not affect the mean square error of our test statistic, which does not hold for the scale transformations. Section 4 discusses the empirical density of the proposed test statistic. Section 5 discusses the critical values of the test statistic. and section 6 discusses the power of the test statistic. Section 7 is devoted to power comparison with some tests proposed earlier. Section 8 illustrates the application of the proposed test on six different real-life data sets and Section 9 concludes and suggests future possible work. The process for determining the critical value, power, empirical density, and p-value of the test statistics is outlined in the appendix and includes fully functional code.

	\section{ Characterization results based on the cumulative and past extropy of record values } 
	The following lemma due to Fashandi and Ahmadi (2012) \cite{fashandiahmadi12} gives a result about symmetric distribution.
	
	\begin{lemma} \label{fa2012}
		(Fashandi and Ahmadi, 2012 \cite{fashandiahmadi12}) Let $X$ be a continuous	random variable with pdf $f_X$ and cdf $F_X$ with support $S_X$. Then, the	identity,  $$f_X(F_X^{-1} (u))=f_X(F_X^{-1} (1-u))$$ for almost all $ u \in (0,\frac{1}{2})$ if and only if  that there exists a constant $k$ such that $F_X(k-x)+F_X(k+x)=1$ for all $x\in S_X.$
	\end{lemma}

	Let $\mathbb{C}$ denote the class of all continuous pdf $f_X$, having cdf $F_X$ such that  $f_X\left(F_X^{-1}(1-u)\right)\geq(\leq)f_X\left(F_X^{-1}(u)\right)$ for all $u\in (0,\frac{1}{2})$. It can be observed using Lemma \ref{fa2012} that $F$ is symmetric if and only if $f_X(F_X^{-1} (u))=f_X(F_X^{-1} (1-u))$ for almost all $ u \in (0,\frac{1}{2})$. The class $\mathbb{C}$  is non-empty and includes but is not limited to power distribution, pareto distribution, exponential distribution, uniform distribution and standard normal distribution (see, Ahmadi (2021) \cite{ahmadi} and Gupta and Chaudhary (2022) \cite{guptaskc}).
	
	Gupta and Chaudhary (2022) \cite{guptaskc} proved the following theorems that will be used to derive test statistics for testing the symmetry of a continuous symmetric distribution.
	\begin{theorem} \label{g} (Gupta and Chaudhary (2022) \cite{guptaskc}) The following two statements are equivalent for any $F_X\in \mathbb{C}:$
		\begin{enumerate}[(i)] \item random variable $X$ has a symmetric distribution;
			\item $\bar{\xi} J(X)=\xi J(X)$. 
		\end{enumerate}  
	\end{theorem} 
	
	\begin{theorem}\label{cpecre} \label{thm2} (Gupta and Chaudhary (2022) \cite{guptaskc}) Let $X_1,X_2,\ldots$ be a random sample of continuous random variables from a population $X$ having cdf $F_X$ and pdf $f_X$.
		The following two statements are equivalent for any $F_X \in \mathbb{C}:$
		\begin{enumerate}[(i)]   	
			\item   random variable $X$ has a symmetric distribution;
			\item for a fixed $k\geq 1$,  $\bar{\xi} J(L_{n,k})=\xi J(U_{n,k})$ for all $n\geq 1$.
		\end{enumerate}
	\end{theorem}
	
	\begin{example} The pdf and cdf respectively of power function distribution are
		\begin{align}
			f_X(x)=\theta x^{\theta -1} \ \ \ \text{and} \ \ \ F_X(x)= x^\theta, \ \  0<x<1, \ \theta >0. 
		\end{align}
		
		Using theorem \ref{g}, Gupta and Chaudhary (2022) \cite{guptaskc} showed  that the power distribution is symmetric for $\theta=1$ .\\
		 
	\end{example}
	\begin{example} The pdf and cdf respectively of Pareto distribution are
		\begin{align}
			f_X(x)=\theta x^{-\theta -1} \ \ \ \text{and} \ \ \ F_X(x)= 1-x^{-\theta}, \ \  x>1 , \ \theta >0. 
		\end{align}

	   Using theorem \ref{cpecre},	Gupta and Chaudhary (2022) \cite{guptaskc} showed that the Pareto distribution is not symmetric.
	\end{example}

	This motivates us that $\boldsymbol{\Delta}_{n,k} =\bar{\xi}  J(L_{n,k})-\xi J(U_{n,k})$ can be used to test whether $X$ is symmetric distribution. Small or large values of $\boldsymbol{\Delta}_{n,k}$ can be regarded as a symptom of non-symmetry and therefore we reject the null hypothesis of symmetry. For this reason, we propose a new test for symmetry based on the sample estimator of $\boldsymbol{\Delta}_{n,k}$ in section \ref{s3}.
	
	\section{ Our proposed test statistics for testing symmetry}\label{s3}
	We see from theorem \ref{cpecre} that  $\boldsymbol{\Delta}_{n,k} =0$ if and only if $X$ is symmetric distribution. Therefore, if an iid sample of size $N$ is available, its empirical counterpart $\boldsymbol{\hat{\Delta}}_{n,k}$ will be helpful in determining whether or not the sample comes from a symmetric distribution. To derive an expression for  $\boldsymbol{\Delta}_{n,k}$, we need to derive an expression for cumulative past extropy of $n$th lower $k$-record value and cumulative residual extropy of $n$th upper $k$-record value.
	
	The cumulative past extropy of $X$ is given as:
	\begin{align*}
		\bar{\xi} J (X) &=- \frac{1}{2}\int_{-\infty}^{\infty} F_{X}^2 (x) dx \nonumber\\
		&= - \frac{1}{2}\int_{0}^{1}  \frac{u^2du}{f(F^{-1}(u))}\\
		&= - \frac{1}{2}\int_{0}^{1} u^2 \left(\frac{dF^{-1} (u)}{du}\right) du.
	\end{align*}
	The cumulative residual extropy  of $X$ is given as :
	\begin{align}
		\xi J (X) &=- \frac{1}{2} \int_{-\infty}^{\infty} \bar{F}_{X}^2 (x) dx \nonumber\\
		&= - \frac{1}{2}\int_{0}^{1}  \frac{(1-u)^2du} {f(F^{-1}(u))}.
	\end{align}
	We propose test statistics as:
	\begin{align}
		\boldsymbol{{\Delta}} &=\xi J (X)-\bar{\xi} J (X)  \nonumber \\
		&= - \frac{1}{2} \int_{0}^{1}  ((1-u)^2 - u^2 ) \frac{du} {f(F^{-1}(u))}\\
		&=  - \frac{1}{2} \int_{0}^{1}  (1-2u )  \left(\frac{dF^{-1} (u)}{du}\right) du \label{teststatistics1}.
	\end{align}
	Following the idea of Vasicek (1976), \cite{vasicek} an estimator of  $\boldsymbol{\Delta}$  will be calculated by replacing the distribution function $F$ by empirical distribution function $\hat{F}_N$ and using the difference operator in place of a differential operator. The derivative of $F^{-1} (u)$ with respect to $u$, that is, $\frac{dF^{-1} (u)}{du}$ will be estimated as 
	\begin{align*}
		\frac{X_{i+m:N}-X_{i-m:N}}{\hat{F}_N(X_{i+m:N})-\hat{F}_N(X_{i-m:N}}=\frac{X_{i+m:N}-X_{i-m:N}}{\frac{i+m}{N}-\frac{i-m}{N}}=\frac{X_{i+m:N}-X_{i-m:N}}{2m/N}.  
	\end{align*}
	Here $m$ denotes window size and it is a positive integer which assume values less than $\frac{N}{2}.$ If $i+m > N$ then we consider $X_{i+m:N} = X_{N:N}$ and if $i+m < 1$ then we consider $X_{i-m:N} = X_{1:N}.$

	Analogous to Vasicek (1976) \cite{vasicek}, Park (1999) \cite{park}, Xiong et al.(2021) \cite{xiongetal}, Jose and Sathar (2022) \cite{josesathar},	we write estimator of $\bar{\xi} J (X)$,  $\xi J (X)$ and  $\boldsymbol{{\Delta}}$ as follows.
	Estimator of $\bar{\xi} J (X)$ is
	\begin{align}
		\hat{D_1}=-\frac{1}{2N} \sum_{i=1}^{N}\left(\frac{i}{N+1}\right)^2 \frac{(X_{i+m:N}-X_{i-m:N})}{2m/N} \nonumber.
	\end{align}
	An estimator of $\xi J (X)$ is
	\begin{align}
		\hat{D_2}=-\frac{1}{2N} \sum_{i=1}^{N}\left(1- \frac{i}{N+1}\right)^2 \frac{(X_{i+m:N}-X_{i-m:N})}{2m/N}.
	\end{align}
	We write estimator of $\boldsymbol{{\Delta}}$ as:
	\begin{align*}
		\boldsymbol{\hat{\Delta}} &=\hat{D_2}-\hat{D_1}\\
		&= - \frac{1}{2N} \sum_{i=1}^{N} \left[1- \frac{2i}{N+1} \right] \frac{(X_{i+m:N}-X_{i-m:N})}{2m/N}.
	\end{align*}
	Cumulative past extropy  of $n$th lower $k$ record value is 
	\begin{align*}
		\bar{\xi} J (L_{n,k}) &=- \frac{1}{2} \int_{-\infty}^{\infty} F_{L_{n,k}}^2 (x) dx \\
		&= - \frac{1}{2} \int_{-\infty}^{\infty} \left(  F^k(x) \sum_{j=0}^{n-1} \frac{(-klog F(x))^j}{j!}\right)^2 dx  \nonumber  \\
		&= - \frac{1}{2} \int_{0}^{1} \left(u^k \sum_{j=0}^{n-1} \frac{(-klog (u))^j }{j!}  \right)^2 \frac{du}{f(F^{-1}(u))}\\
		&= - \frac{1}{2} \int_{0}^{1} u^{2k} \left( \sum_{j=0}^{n-1} \frac{(-klog (u))^j }{j!}  \right)^2 \left(\frac{d}{du} F^{-1}(u) \right)du.
	\end{align*}
	Estimator of $\bar{\xi} J (L_{n,k})$ is
	\begin{align*}
		D^{(1)}_{n,k}&= - \frac{1}{2N} \sum_{i=0}^{N} \left[\left(\frac{i}{N+1}\right)^{2k} \left( \sum_{j=0}^{n-1} \frac{(-klog (\frac{i}{N+1}))^j }{j!}  \right)^2\right. \\
		& \ \ \ \ \ \ \ \ \ \ \ \ \ \ \  \ \ \ \  \ \ \ \ \ \ \ \ \ \ \ \ \ \  \ \ \ \  \left. \left(\frac{X_{i+m:N}-X_{i-m:N}}{2m/N} \right)\right].
	\end{align*}
	Cumulative residual extropy  of $n$th lower $k$ record value is 
	\begin{align}
		\xi J (U_{n,k}) &=- \frac{1}{2} \int_{-\infty}^{\infty} \bar{F}_{U_{n,k}}^2 (x) dx \\
		&= - \frac{1}{2} \int_{-\infty}^{\infty} \left( \bar{F}^k(x) \sum_{j=0}^{n-1} \frac{(-klog \bar{F}(x))^j}{j!} \right)^2 dx  \nonumber \\
		&= - \frac{1}{2} \int_{0}^{1} \left[(1-u)^{2k} \left( \sum_{j=0}^{n-1} \frac{(-klog (1-u))^j  }{j!}  \right)^2 \frac{1}{f(F^{-1}(u))}\right] du \nonumber\\
		&= - \frac{1}{2} \int_{0}^{1} (1-u)^{2k} \left( \sum_{j=0}^{n-1} \frac{(-klog (1-u))^j  }{j!}  \right)^2 \frac{d}{du} (F^{-1}(u))du. \nonumber
	\end{align}
	Estimator of $\bar{\xi} J (U_{n,k})$ is
	\begin{align*}
		D^{(2)}_{n,k}&= - \frac{1}{2N} \sum_{i=0}^{N} \left[\left(1-\frac{i}{N+1}\right)^{2k} \left( \sum_{j=0}^{n-1} \frac{(-klog (1-\frac{i}{N+1}))^j }{j!}  \right)^2 \right. \\ & \ \ \ \ \ \ \ \ \ \ \ \  \ \ \ \ \ \ \ \ \ \ \ \  \  \ \ \ \ \ \ \ \ \ \ \ \ \left. \left(\frac{X_{i+m:N}-X_{i-m:N}}{2m/N} \right)\right].
	\end{align*}
	We propose test statistics as:
	\begin{align}
		\boldsymbol{{\Delta}}_{n,k}&=\xi J (U_{n,k})-\bar{\xi} J (L_{n,k})  \nonumber \\
		&= - \frac{1}{2} \int_{0}^{1} (1-u)^{2k}\left( \sum_{j=0}^{n-1} \frac{(-klog (1-u))^j  }{j!}  \right)^2 \frac{du}{f(F^{-1}(u)}  \nonumber\\ & \ \ \ \ \ \ \ \ \ \ \ \ \ \ \ + \frac{1}{2} \int_{0}^{1} u^{2k} \left( \sum_{j=0}^{n-1} \frac{(-klog (u))^j }{j!}  \right)^2 \frac{du}{f(F^{-1}(u)} \nonumber\\
		&= - \frac{1}{2} \int_{0}^{1}  \left[ (1-u)^{2k} \left( \sum_{j=0}^{n-1} \frac{(-klog (1-u))^j }{j!}  \right)^2\right.  \nonumber \\ & \ \ \ \ \ \ \ \ \  \ \ \ \ \ \ \ \   \ \ \ \ \ \ \ \  \ \ \left. - u^{2k} \left( \sum_{j=0}^{n-1} \frac{(-klog (u))^j }{j!}  \right)^2   \right] \frac{du}{f(F^{-1}(u)} \nonumber\\
		&=  - \frac{1}{2} \int_{0}^{1}  \left[(1-u)^{2k} \left( \sum_{j=0}^{n-1} \frac{(-klog (1-u))^j }{j!}  \right)^2 \right.  \nonumber \\ & \ \ \ \ \ \ \ \ \  \ \ \ \ \ \ \ \   \ \ \ \ \ \ \ \  \ \ \left. - u^{2k} \left( \sum_{j=0}^{n-1} \frac{(-klog (u))^j u^k}{j!}  \right)^2   \right] \left(\frac{dF^{-1} (u)}{du}\right) du .\nonumber
	\end{align}
	An estimator of test statistics $\boldsymbol{{\Delta}}_{n,k}$ is 
	\begin{align}
		&\boldsymbol{\hat{\Delta}}_{n,k} =D^{(2)}_{n,k}-D^{(1)}_{n,k} \nonumber\\
		&= - \frac{1}{2N} \sum_{i=1}^{N} \left[\left(1-\frac{i}{N+1}\right)^{2k} \left( \sum_{j=0}^{n-1} \frac{(-klog (1-\frac{i}{N+1}))^j  }{j!}  \right)^2 \right. \nonumber\\ &\  \left.-\left(\frac{i}{N+1}\right)^{2k} \left( \sum_{j=0}^{n-1} \frac{(-klog (\frac{i}{N+1}))^j }{j!}  \right)^2   \right] \frac{(X_{i+m:N}-X_{i-m:N})}{2m/N} .\nonumber 
	\end{align}		
	We choose $n=2$ and $k=2$ for the simplicity of calculation, so we consider test statistics as $\hat{\Delta}_{2,2} $. 
	\begin{align}
		\boldsymbol{\hat{\Delta}}_{2,2} &= - \frac{1}{2N} \sum_{i=1}^{N} \left[\left(1-\frac{i}{N+1}\right)^{4} \left(1-2log (1-\frac{i}{N+1})  \right)^2\right.\nonumber\\ & \ \ \ \ \  \left. -\left(\frac{i}{N+1}\right)^{4} \left( 1-2log (\frac{i}{N+1})\right)^2   \right] \frac{(X_{i+m:N}-X_{i-m:N})}{2m/N}. \nonumber
	\end{align}		
	
	The procedure remains the same we choose any other $n$ and $k$. Following theorem says $\boldsymbol{\hat{\Delta}}_{2,2}$ is consistent estimator of $\boldsymbol{\Delta_{2,2}}.$ 
	\begin{theorem}
		Assume that $X_1,\ X_2,\,..., X_N$ is a random sample of size $N$ taken from a population with pdf $f$ and cdf $F$. Also, let the variance of the random variable be finite. Then $\boldsymbol{\hat{\Delta}}_{2,2}$ converges in probability to $\Delta_{2,2}$ as $N \longrightarrow \infty, \ m\longrightarrow \infty \ \text{and} \ \frac{m}{N} \longrightarrow 0.$
	\end{theorem}
	\textbf{Proof : } Following lines of the proof of Theorem 1 of Vasicek (1976) \cite{vasicek}. We have, $D^{(2)}_{2,2} \overset{P}{\to} \xi J (U_{2,2})$ and  $D^{(1)}_{2,2} \overset{P}{\to} \bar{\xi}J (L_{2,2})$. Therefore we get,  $\boldsymbol{\hat{\Delta}}_{2,2} \overset{P}{\to}  \boldsymbol{\Delta}_{2,2}$. That is, $\boldsymbol{\hat{\Delta}}_{2,2}$ is consistent estimator of  $\boldsymbol{\Delta}_{2,2}$.\\
	\indent Note that the test statistics proposed by Park (1999) \cite{park}, Xiong et al. (2021) \cite{xiongetal} and  Jose and Sathar (2022) \cite{josesathar} are consistent due to the method given in Vasicek (1976) \cite{vasicek}.
	
	\begin{theorem}
		Let $X_1, X_2, ... , X_N$ be a sequence of iid random variables and let $Y_i=aX_i+b, \ a>0,\ b\in \mathbb{R},\ i=1,2,... ,N.$ Denote the estimator for $\Delta_{2,2}$ based on $X_i$ and $Y_i$ by $\Delta_{2,2}^X$ and $\Delta_{2,2}^Y$, respectively. Then 
		\begin{enumerate}[(i)] 
			\item  E($\Delta_{2,2}^Y)=a E(\Delta_{2,2}^X)$
			\item  Var($\Delta_{2,2}^Y)=a^2 Var(\Delta_{2,2}^X)$
			\item  MSE($\Delta_{2,2}^Y)=a^2 MSE(\Delta_{2,2}^X)$\\
		\end{enumerate}
		where $E(X),\ Var(X)$ and $MSE(X)$ represent expectation, variance and mean square error of random variable $X$, respectively.
	\end{theorem}
	\textbf{Proof: }
	\begin{align}
		\newline   \boldsymbol{\hat{\Delta}}_{2,2}^Y 
		= & - \frac{1}{2N} \sum_{i=1}^{N} \left[\left(1-\frac{i}{N+1}\right)^{4}  \left(1-2log ( 1 - \frac{i}{N+1} ) \right)^2 \right.\nonumber \\ & \left.-\left(\frac{i}{N+1}\right)^{4} \left( 1-2log (\frac{i}{N+1})\right)^2   \right] \frac{(Y_{i+m:N}-Y_{i-m:N})}{2m/N} \nonumber\\
		= & - \frac{1}{2N} \sum_{i=1}^{N} \left[\left(1-\frac{i}{N+1}\right)^{4} \left(1-2log (1-\frac{i}{N+1})  \right)^2  \right.\nonumber \\ & \left. -\left(\frac{i}{N+1}\right)^{4} \left( 1-2log (\frac{i}{N+1})\right)^2   \right] \frac{(aX_{i+m:N}-aX_{i-m:N})}{2m/N} \nonumber\\
		= & a\boldsymbol{\hat{\Delta}}_{2,2}^X .\nonumber
	\end{align}
	Thus, proof is completed because of $\boldsymbol{\hat{\Delta}}_{2,2}^Y=a\boldsymbol{\hat{\Delta}}_{2,2}^X$ and  properties of mean, variance and MSE of $X$.
	
	\section{Empirical density of $\boldsymbol{\hat{\Delta}}_{2,2}$}
	
	For the computation of the critical values, it is crucial to assess the asymptotic distribution of the newly suggested test statistic. Unfortunately, it is very difficult to deduce the distribution of $\boldsymbol{\hat{\Delta}}_{2,2}$ as $N\rightarrow$ $\infty$, because there is a window size $m$ that depends on $N$. In light of this, the empirical density of the suggested test statistic is evaluated. Figure 1 depicts the empirical densities of $\boldsymbol{\hat{\Delta}}_{2,2}$ for sample size $N=100$ and window size $m=40$ based on $10,000$ iterations generated from the null distribution taken as standard normal distribution.

	\begin{center}
		{\bf Figure 1}. \small{{Empirical densities of $\boldsymbol{\hat{\Delta}}_{2,2}$ for $N=100$ and $m=40$}} 
	\end{center}
	\includegraphics{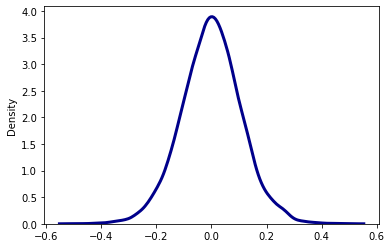}\label{figure1}

	\section{ Critical values of $|\boldsymbol{\hat{\Delta}}_{2,2}|$ }
	
	Now, 10,000 samples of size $N$ taken from the null distribution, which serves as the standard normal distribution, are used to calculate the critical values of the $\boldsymbol{\hat{\Delta}}_{2,2}$ function. Then, the values of $\boldsymbol{\hat{\Delta}}_{2,2}$ are obtained based on these 10,000 samples of size $N$ generated from the null distribution. From the 10,000 values of  $\boldsymbol{\hat{\Delta}}_{2,2}$, $(1-\frac{\alpha}{2})$th quantile represents the critical value corresponding to sample size $N$ of the test statistic at significance level $\alpha$. If the critical values are denoted as  $\boldsymbol{\hat{\Delta}}_{2,2}(1-\frac{\alpha}{2})$, then the null hypothesis is rejected with size $\alpha$ whenever $|\boldsymbol{\hat{\Delta}}_{2,2}|>\boldsymbol{\hat{\Delta}}_{2,2}(1-\frac{\alpha}{2}).$

	The exact critical values of $\boldsymbol{\hat{\Delta}}_{2,2}$ based on 10,000 samples of different sizes generated from the null distribution at significance level $\alpha=0.10$, $\alpha=0.05$ and $\alpha=0.01$ are given in Table 1, 2 and 3 respectively. For sample sizes $N=5,10,20,30,40,50,100$ with window sizes $m$ ranging from 1 to 30, the critical values are obtained. The next section deals with the simulation study through which the power of the test statistic is evaluated. 
	
	\begin{center}
		{\bf Table 1}. \small{{Critical valus of $|\boldsymbol{\hat{\Delta}}_{2,2}|$ statistics at significance level $\alpha$= 0.10}} 
		
		\resizebox{!}{!}{
			\begin{tabular}{ p{1.0cm} p{1.0cm} p{1.0cm} p{1.0cm} p{1.0cm} p{1.0cm} p{1.0cm} p{1.0cm} } 
				\hline
				$m\backslash N$ & 5 & 10 & 20  & 30  & 40 & 50 & 100 \\
				\hline \\
				2  & 0.3151   & 0.5250 & 0.5743 & 0.5777 & 0.5738 & 0.5590 & 0.5197 \\
				3 & \   & 0.4141  & 0.5179  & 0.5092 & 0.5176 & 0.5071 & 0.4824 \\ 
				4 & \   & 0.3167  & 0.4581  & 0.4783 & 0.4818 & 0.4762 & 0.4572 \\
				5 & \   & \  & 0.4183  & 0.4501 & 0.4535 & 0.4588 &  0.4401 \\
				6 & \   & \  & 0.3778  & 0.4256 & 0.4436 &  0.4360 & 0.4187 \\
				7 & \   & \  & 0.3330  & 0.3907 & 0.4131 & 0.4148 & 0.4149 \\
				8 & \   & \  & 0.2948  & 0.3652 & 0.3901 & 0.4076 & 0.3955 \\
				9  & \   & \  & 0.2584  & 0.3464 & 0.3747 & 0.3929 & 0.3874 \\
				10 & \   & \  & \  & 0.3190 & 0.3621 & 0.3735 & 0.3862 \\
				11 & \   & \  & \  & 0.2989 & 0.3409 & 0.3681 & 0.3676 \\
				12 & \   & \  & \  & 0.2808 & 0.3291 & 0.3483 & 0.3682 \\
				13 & \   & \  & \  & 0.2542 & 0.3099 & 0.3428 & 0.3672 \\
				14 & \   & \  & \  & 0.2324 & 0.2991 & 0.3311 & 0.3631 \\
				15 & \   & \  & \  & \ & 0.2814 & 0.3150 & 0.3528 \\
				16 & \   & \  & \  & \ & 0.2667 & 0.3075 & 0.3442 \\
				17 & \   & \  & \  & \ & 0.2497 & 0.2906 & 0.3373 \\
				18 & \   & \  & \  & \ & 0.2379 & 0.2868 & 0.3358 \\
				19 & \   & \  & \  & \ & 0.2251 & 0.2738 & 0.3375 \\
				20 & \   & \  & \  & \ & \ & 0.2589 & 0.3329 \\
				21 & \   & \  & \  & \ & \ & 0.2500 & 0.3205 \\
				22 & \   & \  & \  & \ & \ & 0.2395 & 0.3163 \\
				23 & \   & \  & \  & \ & \ & 0.2305 & 0.3126 \\
				24 & \   & \  & \  & \ & \ & 0.2212 & 0.3097 \\
				25 & \   & \  & \  & \ & \ & \ & 0.3111 \\
				26 & \   & \  & \  & \ & \ & \ & 0.2977 \\
				27 & \   & \  & \  & \ & \ & \ & 0.2957 \\
				28 & \   & \  & \  & \ & \ & \ & 0.2972 \\
				29 & \   & \  & \  & \ & \ & \ & 0.2896 \\
				30 & \   & \  & \  & \ & \ & \ & 0.2827 \\
				40 & \   & \  & \  & \ & \ & \ & 0.2409 \\
				
				\hline
		\end{tabular}}\label{table1}
	\end{center}

	\begin{center}
		{\bf Table 2}.  \small{{Critical valus of $|\boldsymbol{\hat{\Delta}}_{2,2}|$ statistics at significance level $\alpha$= 0.05}} 
		
		\resizebox{!}{!}{
			\begin{tabular} {p{1.0cm} p{1.0cm} p{1.0cm} p{1.0cm} p{1.0cm} p{1.0cm} p{1.0cm} p{1.0cm} p{1.0cm} p{1.0cm} p{1.0cm} } 
				\hline
				$m\backslash N$ & 5 & 10 & 20  & 30  & 40 & 50 & 100 \\
				\hline 
				\vspace{0.1cm}\\
				2 & 0.3637 & 0.6093  & 0.6673  & 0.6703 & 0.6658 & 0.6474 & 0.5969 \\
				3 & \   & 0.4787  & 0.5833  & 0.5936 & 0.6011 & 0.5857 & 0.5611 \\ 
				4 & \   & 0.3641  & 0.5333  & 0.5539 & 0.5553 & 0.5387 & 0.5284 \\
				5 & \   & \  & 0.4776 & 0.5216 & 0.5287 & 0.5305 & 0.5054 \\
				6 & \   & \  & 0.4362  & 0.4872 & 0.5074 & 0.4979 & 0.4794 \\
				7 & \   & \  & 0.3848  & 0.4536 & 0.4785 & 0.4718 & 0.4750 \\
				8 & \   & \  & 0.3460  & 0.4207 & 0.4543 & 0.4647 & 0.4573 \\
				9  & \   & \  & 0.2951  & 0.4044 & 0.4271 & 0.4491 & 0.4488 \\
				10 & \   & \  & \  & 0.3642 &  0.4188 & 0.4235 & 0.4405 \\
				11 & \   & \  & \  & 0.3440 & 0.3953 & 0.4259 & 0.4242 \\
				12 & \   & \  & \  & 0.3254 & 0.3775 & 0.4053 & 0.4249 \\
				13 & \   & \  & \  & 0.2948 & 0.3561 & 0.3947 & 0.4251 \\
				14 & \   & \  & \  & 0.2751 & 0.3515 & 0.3821 & 0.4203 \\
				15 & \   & \  & \  & \ & 0.3239 & 0.3626 & 0.4157 \\
				16 & \   & \  & \  & \ & 0.3057 & 0.3521 & 0.3967  \\
				17 & \   & \  & \  & \ & 0.2868 & 0.3433 & 0.3901 \\
				18 & \   & \  & \  & \ & 0.2769 & 0.3356 & 0.3820 \\
				19 & \   & \  & \  & \ & 0.2583 & 0.3182 & 0.3883 \\
				20 & \   & \  & \  & \ & \ & 0.3011 & 0.3779 \\
				21 & \   & \  & \  & \ & \ & 0.2909 & 0.3646 \\
				22 & \   & \  & \  & \ & \ & 0.2777 & 0.3705 \\
				23 & \   & \  & \  & \ & \ & 0.2697 & 0.3607 \\
				24 & \   & \  & \  & \ & \ & 0.2533 & 0.3575 \\
				25 & \   & \  & \  & \ & \ & \ & 0.3592 \\
				26 & \   & \  & \  & \ & \ & \ & 0.3446 \\
				27 & \   & \  & \  & \ & \ & \ & 0.3358 \\
				28 & \   & \  & \  & \ & \ & \ & 0.3427 \\
				29 & \   & \  & \  & \ & \ & \ & 0.3355 \\
				30 & \   & \  & \  & \ & \ & \ & 0.3258 \\
				40 & \   & \  & \  & \ & \ & \ & 0.2797 \\

				\hline
		\end{tabular}}
	\end{center}\label{table2}

	\begin{center}
		{\bf Table 3}.  \small{{Critical valus of $|\boldsymbol{\hat{\Delta}}_{2,2}|$ statistics at significance level $\alpha$= 0.01}} 
		
		\vspace{0.2cm}
		\resizebox{!}{!}{
			\begin{tabular}{ p{1.0cm} p{1.0cm} p{1.0cm} p{1.0cm} p{1.0cm} p{1.0cm} p{1.0cm} p{1.0cm} } 
				\hline
				$m\backslash N$ & 5 & 10 & 20  & 30  & 40 & 50 & 100 \\
				\hline \\
				2  & 0.4569   & 0.7690  & 0.8663  & 0.8474 & 0.8501 & 0.8292 & 0.7548 \\
				3 & \   & 0.6042  & 0.7349  & 0.7590 & 0.7570 & 0.7447 & 0.6997 \\ 
				4 & \   & 0.4739  & 0.6810  & 0.7185 & 0.6946 & 0.6843 & 0.6553 \\
				5 & \   & \  & 0.6158  & 0.6570 & 0.6695 & 0.6629 & 0.6488 \\
				6 & \   & \  & 0.5636  & 0.6427 & 0.6256 & 0.6502 & 0.6257 \\
				7 & \   & \  & 0.4903  & 0.5819 & 0.6082 & 0.5989 & 0.6017 \\
				8 & \   & \  & 0.4428  & 0.5273 & 0.5784 & 0.6020 & 0.5835 \\
				9  & \   & \  & 0.3757  & 0.5063 & 0.5343 & 0.5828 & 0.5844 \\
				10 & \   & \  & \  & 0.4594 & 0.5488 & 0.5394 & 0.5560 \\
				11 & \   & \  & \  & 0.4536 & 0.4971 & 0.5462 & 0.5438 \\
				12 & \   & \  & \  & 0.4146 & 0.4845  & 0.5288 & 0.5504 \\
				13 & \   & \  & \  & 0.3862 & 0.4462 & 0.5049 & 0.5480 \\
				14 & \   & \  & \  & 0.3467 & 0.4364 & 0.4878 & 0.5433 \\
				15 & \   & \  & \  & \ & 0.4213 & 0.4523 & 0.5259 \\
				16 & \   & \  & \  & \ & 0.4091 & 0.4747 & 0.5016 \\
				17 & \   & \  & \  & \ & 0.3727 & 0.4471 & 0.5187 \\
				18 & \   & \  & \  & \ & 0.3572 & 0.4239 & 0.4955 \\
				19 & \   & \  & \  & \ & 0.3285 & 0.4134 & 0.4912 \\
				20 & \   & \  & \  & \ & \ & 0.3869 & 0.4794 \\
				21 & \   & \  & \  & \ & \ & 0.3777 & 0.4761 \\
				22 & \   & \  & \  & \ & \ & 0.3587 & 0.4931 \\
				23 & \   & \  & \  & \ & \ & 0.3505 & 0.4606 \\
				24 & \   & \  & \  & \ & \ & 0.3241 & 0.4648 \\
				25 & \   & \  & \  & \ & \ & \ & 0.4524 \\
				26 & \   & \  & \  & \ & \ & \ & 0.4469 \\
				27 & \   & \  & \  & \ & \ & \ & 0.4388 \\
				28 & \   & \  & \  & \ & \ & \ & 0.4405 \\
				29 & \   & \  & \  & \ & \ & \ & 0.4270 \\
				30 & \   & \  & \  & \ & \ & \ & 0.4086 \\
				40 & \   & \  & \  & \ & \ & \ & 0.3650 \\
				
				\hline
		\end{tabular}}\label{table3}
	\end{center}

	\section{ Power of  $\boldsymbol{\hat{\Delta}}_{2,2}$ }
	Unfortunately, it is very complicated to derive the exact distribution of $\boldsymbol{\hat{\Delta}}_{2,2}$ because it depends on window size $m$ which is dependent on sample size $N$. Tables 1, 2 and 3 show the exact critical values of the test statistic $|\boldsymbol{\hat{\Delta}}_{2,2}|$ for various sample sizes by Monte Carlo simulation with 10,000 repetitions, for significance levels $\alpha= 0.10$, $\alpha = 0.05$, and $\alpha = 0.01.$ A similar procedure has been used in Xiong et al. (2021)\cite{xiongetal}. In order to determine whether the test statistic's absolute value is greater than the critical value, we generated a sample of size $N$ from the null distribution. We then repeated this process 10,000 times. The power of the test is measured by the percentage of rejection. Table 4,5 and 6 gives power of test when alternative distribution is taken as chi-square distribution with degree of freedom 1 for $\alpha= 0.10$,  $\alpha = 0.05$ and $\alpha = 0.01$, respectively. Since $\chi^2(1)$ is asymmetric distribution. Our test significantly verifies this fact. We observe from table 4,5,6 that when we increase sample size $N$, power increases. When the sample size is 100, power is 1.000 for any value of $m$. That means our test performs well for a large sample size.

	\begin{center}
		{\bf Table 4}.  \small{{Powers of  $\boldsymbol{\hat{\Delta}}_{2,2}$ statistics against alternative $\chi^2_{(1)}$ at significance level $\alpha= 0.10$}}
		
		\vspace{0.2cm}
		\resizebox{!}{!}{
			\begin{tabular}{ p{1.0cm} p{1.0cm} p{1.0cm} p{1.0cm} p{1.0cm} p{1.0cm} p{1.0cm} p{1.0cm} } 
				\hline
				$m\backslash N$ & 5 & 10 & 20  & 30  & 40 & 50 & 100 \\
				\hline \\
				2  & 0.3168   & 0.6397  & 0.9151  & 0.9822 & 0.9966 & 0.9998 & 1.0000 \\
				3 & \   & 0.6285  & 0.9186  & 0.9870 & 0.9971 & 0.9996 & 1.0000 \\ 
				4 & \   & 0.6127  & 0.9220  & 0.9885 & 0.9976 & 0.9999 & 1.0000 \\
				5 & \   & \  & 0.9166  & 0.9856 & 0.9990 & 0.9999 & 1.0000 \\
				6 & \   & \  & 0.9171  & 0.9848 & 0.9970 & 0.9996 & 1.0000 \\
				7 & \   & \  & 0.9125  & 0.9851 & 0.9979 & 0.9998 & 1.0000 \\
				8 & \   & \  & 0.9106  & 0.9850 & 0.9984 & 0.9996 & 1.0000 \\
				9  & \   & \  & 0.8980  & 0.9841 & 0.9981 & 0.9996 & 1.0000 \\
				10 & \   & \  & \  & 0.9846 & 0.9970 & 0.9998 & 1.0000 \\
				11 & \   & \  & \  & 0.9836 & 0.9978 & 0.9993 & 1.0000 \\
				12 & \   & \  & \  & 0.9804 & 0.9984 & 1.0000 & 1.0000 \\
				13 & \   & \  & \  & 0.9763 & 0.9972 & 0.9995 & 1.0000 \\
				14 & \   & \  & \  & 0.9776 & 0.9969 & 0.9996 & 1.0000 \\
				15 & \   & \  & \  & \ & 0.9963 & 0.9997 & 1.0000 \\
				16 & \   & \  & \  & \ & 0.9957 & 0.9994 & 1.0000 \\
				17 & \   & \  & \  & \ & 0.9960 & 0.9996 & 1.0000 \\
				18 & \   & \  & \  & \ & 0.9952 & 0.9991 & 1.0000 \\
				19 & \   & \  & \  & \ & 0.9948 & 0.9992 & 1.0000 \\
				20 & \   & \  & \  & \ & \ & 0.9994 & 1.0000 \\
				21 & \   & \  & \  & \ & \ & 0.9987 & 1.0000 \\
				22 & \   & \  & \  & \ & \ & 0.9992 & 1.0000 \\
				23 & \   & \  & \  & \ & \ & 0.9987 & 1.0000 \\
				24 & \   & \  & \  & \ & \ & 0.9982 & 1.0000 \\
				25 & \   & \  & \  & \ & \ & \ & 1.0000 \\
				26 & \   & \  & \  & \ & \ & \ & 1.0000 \\
				27 & \   & \  & \  & \ & \ & \ & 1.0000 \\
				28 & \   & \  & \  & \ & \ & \ & 1.0000 \\
				29 & \   & \  & \  & \ & \ & \ & 1.0000 \\
				30 & \   & \  & \  & \ & \ & \ & 1.0000 \\
				40 & \   & \  & \  & \ & \ & \ & 1.0000 \\
				
				\hline
		\end{tabular}}
	\end{center}\label{table4}

	\begin{center}
		\noindent {\bf Table 5}.  \small{{Powers of  $\boldsymbol{\hat{\Delta}}_{2,2}$ statistics against alternative $\chi^2_{(1)}$ at significance level $\alpha$= 0.05}} 
		
		\vspace{0.2cm}
		\resizebox{!}{!}{
			\begin{tabular}{ p{1.0cm} p{1.0cm} p{1.0cm} p{1.0cm} p{1.0cm} p{1.0cm} p{1.0cm} p{1.0cm} } 
				\hline
				$m\backslash N$ & 5 & 10 & 20  & 30  & 40 & 50 & 100 \\
				\hline \\
				2  & 0.2649   & 0.5644  & 0.8759  & 0.9685 & 0.9924 & 0.9984 & 1.0000 \\
				3 & \   & 0.5624  & 0.8859  & 0.9703 & 0.9922 & 0.9991 & 1.0000 \\ 
				4 & \   & 0.5534  & 0.8769  & 0.9749 & 0.9957 & 0.9997 & 1.0000 \\
				5 & \   & \  & 0.8794  & 0.9694 & 0.9952 & 0.9990 & 1.0000 \\
				6 & \   & \  & 0.8691  & 0.9784 & 0.9941 & 0.9991 & 1.0000 \\
				7 & \   & \  & 0.8756  & 0.9756 & 0.9945 & 0.9995 & 1.0000 \\
				8 & \   & \  & 0.8536  & 0.9759 & 0.9952 & 0.9989 & 1.0000 \\
				9  & \   & \  & 0.8556  & 0.9680 & 0.9949 & 0.9993 & 1.0000 \\
				10 & \   & \  & \  & 0.9679 & 0.9936 & 0.9991 & 1.0000 \\
				11 & \   & \  & \  & 0.9708 & 0.9937 & 0.9986 & 1.0000 \\
				12 & \   & \  & \  & 0.9609 & 0.9928 & 0.9991 & 1.0000 \\
				13 & \   & \  & \  & 0.9589 & 0.9945 & 0.9991 & 1.0000 \\
				14 & \   & \  & \  & 0.9586 & 0.9913 & 0.9989 & 1.0000 \\
				15 & \   & \  & \  & \ & 0.9924 & 0.9991 & 1.0000 \\
				16 & \   & \  & \  & \ & 0.9902 & 0.9988 & 1.0000 \\
				17 & \   & \  & \  & \ & 0.9902 & 0.9978 & 1.0000 \\
				18 & \   & \  & \  & \ & 0.9903 & 0.9972 & 1.0000 \\
				19 & \   & \  & \  & \ & 0.9870 & 0.9984 & 1.0000 \\
				20 & \   & \  & \  & \ & \ & 0.9987 & 1.0000 \\
				21 & \   & \  & \  & \ & \ & 0.9977 & 1.0000 \\
				22 & \   & \  & \  & \ & \ & 0.9971 & 1.0000 \\
				23 & \   & \  & \  & \ & \ & 0.9964 & 1.0000 \\
				24 & \   & \  & \  & \ & \ & 0.9958 & 1.0000 \\
				25 & \   & \  & \  & \ & \ & \ & 1.0000 \\
				26 & \   & \  & \  & \ & \ & \ & 1.0000 \\
				27 & \   & \  & \  & \ & \ & \ & 1.0000 \\
				28 & \   & \  & \  & \ & \ & \ & 1.0000 \\
				29 & \   & \  & \  & \ & \ & \ & 1.0000 \\
				30 & \   & \  & \  & \ & \ & \ & 1.0000 \\
				40 & \   & \  & \  & \ & \ & \ & 1.0000 \\
				
				\hline
		\end{tabular}}
	\end{center}\label{table5}

	\begin{center}
		\noindent{\bf Table 6}.  \small{{Powers of  $\boldsymbol{\hat{\Delta}}_{2,2}$ statistics against alternative $\chi^2_{(1)}$ at significance level $\alpha$= 0.01}} 
		
		\resizebox{!}{!}{
			\begin{tabular}{ p{1.0cm} p{1.0cm} p{1.0cm} p{1.0cm} p{1.0cm} p{1.0cm} p{1.0cm} p{1.0cm} } 
				\hline
				$m\backslash N$ & 5 & 10 & 20  & 30  & 40 & 50 & 100 \\
				\hline \\
				2 & 0.1924   & 0.4437  & 0.8671  & 	0.9162 & 0.9714 & 0.9978 & 1.0000 \\
				3 & \   & 0.4483  & 0.7861  & 0.9273 & 0.9760 & 0.9981 & 1.0000 \\
				4 & \   & 0.4114  & 0.7799  & 0.9215 & 0.9825 & 0.9989 & 1.0000 \\
				5 & \   & \  & 0.7713  & 0.9318 & 0.9789 & 0.9991 & 1.0000 \\
				6 & \   & \  & 0.7566  & 0.9186 & 0.9830 & 0.9991 & 1.0000 \\
				7 & \   & \  & 0.7611  & 0.9264 & 0.9818 & 0.9994 & 1.0000 \\
				8 & \   & \  & 0.7496  & 0.9398 & 0.9800 & 0.9991 & 1.0000 \\
				9  & \   & \  & 0.7502  & 0.9302 & 0.9802 & 0.9991 & 1.0000 \\
				10 & \   & \  & \  & 0.9283 & 0.9728 & 0.9990 & 1.0000 \\
				11 & \   & \  & \  & 0.9104 & 0.9813 & 0.9992 & 1.0000 \\
				12 & \   & \  & \  & 0.9127 & 0.9744 & 0.9989 & 1.0000 \\
				13 & \   & \  & \  & 0.8990 & 0.9795 & 0.9933 & 1.0000 \\
				14 & \   & \  & \  & 0.8929 & 0.9738 & 0.9918 & 1.0000 \\
				15 & \   & \  & \  & \ & 0.9721 & 0.9942 & 1.0000 \\
				16 & \   & \  & \  & \ & 0.9611 & 0.9900 & 1.0000 \\
				17 & \   & \  & \  & \ & 0.9649 & 0.9904 & 1.0000 \\
				18 & \   & \  & \  & \ & 0.9565 & 0.9897 & 1.0000 \\
				19 & \   & \  & \  & \ & 0.9588 & 0.9887 & 1.0000 \\
				20 & \   & \  & \  & \ & \ & 0.9979 & 1.0000 \\
				21 & \   & \  & \  & \ & \ & 0.9984 & 1.0000 \\
				22 & \   & \  & \  & \ & \ & 0.9972 & 1.0000 \\
				23 & \   & \  & \  & \ & \ & 0.9951 & 1.0000 \\
				24 & \   & \  & \  & \ & \ & 0.9974 & 1.0000 \\
				25 & \   & \  & \  & \ & \ & \ & 1.0000 \\
				26 & \   & \  & \  & \ & \ & \ & 1.0000 \\
				27 & \   & \  & \  & \ & \ & \ & 1.0000 \\
				28 & \   & \  & \  & \ & \ & \ & 1.0000 \\
				29 & \   & \  & \  & \ & \ & \ & 1.0000 \\
				30 & \   & \  & \  & \ & \ & \ & 1.0000 \\
				40 & \   & \  & \  & \ & \ & \ & 1.0000 \\
				
				\hline
		\end{tabular}}
	\end{center} \label{table6}
	Table 7 gives information about power against different alternative distribution $\chi^2{(1)}$, $\chi^2{(2)}$, $\chi^2{(3)}$ and $N(0,1)$.	\begin{center}
		\noindent{\bf Table 7}.  \small{{Power of $\boldsymbol{\hat{\Delta}}_{2,2}$ for $N=20,50,100$ and significance level $\alpha= 0.05$ for alternatives  $\chi^2_{(1)}$,  $\chi^2_{(2)}$, $\chi^2_{(3)}$  and $N(0,1)$ }} 
		
		\resizebox{!}{!}{
			\begin{tabular}{ p{1.0cm} p{1.0cm} p{1.0cm} p{1.0cm} p{1.0cm}  p{1.0cm}} 
				\hline
				N & m  & $\chi^2_{(1)}$ & $\chi^2_{(2)}$ & $\chi^2_{(3)}$ & $N(0,1)$ \\
				\hline 
				
				\ & 2   & 0.8759  & 0.8861  & 0.8627 & 0.0231  \\
				\ & 3   & 0.8859  & 0.8976 & 0.8728 & 0.0235  \\
				\ & 4   & 0.8769  & 0.8956 & 0.8762 & 0.0224  \\
				\ & 5   & 0.8794  & 0.8943 & 0.8715 & 0.0263  \\
				20 & 6   & 0.8691  & 0.8941 & 0.8732 & 0.0244 \\
				\ & 7  & 0.8756  & 0.8889 & 0.8716 & 0.0253   \\
				\ & 8   & 0.8536  & 0.8814 & 0.8643 & 0.0230   \\
				\ & 9   & 0.8556  & 0.8757 &0.8555 & 0.0210   \\

				\hline
				
				\ & 2   & 0.9984  & 0.9981 & 0.9955 & 0.0222   \\
				\ & 4   & 0.9997  & 0.9991 & 0.9969 & 0.0263  \\
				\ & 7   & 0.9995  & 0.9987 & 0.9964  & 0.0304   \\
				\ & 9   & 0.9993  & 0.9986 & 0.9960 & 0.0218  \\
				50 & 15   &  0.9991  & 0.9985 & 0.9961 & 0.0261  \\
				\ & 17  & 0.9978  & 0.9986 & 0.9955 & 0.0259  \\
				\ & 20  &  0.9987  & 0.9979 & 0.9944 & 0.0226  \\
				\ & 22  & 0.9971  & 0.9976 & 0.9946 & 0.0247  \\

				\hline
				
				\ & 2   & 1.0000  & 1.0000 & 1.0000 & 0.0263   \\
				\ & 4   & 1.0000  & 1.0000 & 1.0000 & 0.0257  \\
				\ & 5   & 1.0000  & 1.0000 & 1.0000 & 0.0238  \\
				\ & 7   & 1.0000  & 1.0000 & 1.0000 & 0.0250  \\
				100 & 10   & 1.0000  & 1.0000 & 1.0000 & 0.0307   \\
				\ & 15 & 1.0000  & 1.0000 &1.0000 & 0.0239   \\
				\ & 20   & 1.0000  & 1.0000 & 1.0000 & 0.0231   \\
				\ & 30  & 1.0000  & 1.0000 & 1.0000  & 0.0225   \\
				\ & 40   & 1.0000  & 1.0000 & 1.0000 & 0.0250   \\
				
				\hline
		\end{tabular}}
	\end{center}\label{table7}
	
	\section{Power comparision}
	We recall that Park (1999) \cite{park}, Xiong et al.(2021) \cite{xiongetal} and Jose and Sathar (2022) \cite{josesathar} also used the idea in Vasicek (1976) and proposed a test for symmetry based on the entropy of order statistics, extropy of $k$th upper and lower record value and extropy of $n$th upper and lower $k$-record value, respectively. Let $T_1$ and $T_2$ denotes test statistics of Xiong et al. (2021) \cite{xiongetal} and Jose and Sathar (2022) \cite{josesathar} respectively. Power of $T_1$ and $T_2$ in table \ref{table8} are taken from respective papers. Park (1999) \cite{park} showed that his test outperformed others. Xiong et al. (2021) compared their test with Park (1999) \cite{park}. If the alternative distribution is  $\chi^2_{(1)}$, $\chi^2_{(2)}$ or $N(0,1)$ then Xiong et al. (2021) \cite{xiongetal} test and Park (1999) \cite{park} test have the almost same performance for moderate and large sample size. 
	Table 8 compares the power of our proposed test with the test proposed by Xiong et al. (2021) \cite{xiongetal} and Jose and Sathar (2022) \cite{josesathar}. Since $\chi^2_{(2)}$ is not symmetric and higher value of power yield a better test. Our test performs better in power comparison than the test proposed by Xiong et al. (2021) \cite{xiongetal} and Jose and Sathar (2022) \cite{josesathar} except in very few cases. Since $N(0,1)$ is symmetric distribution and a lower value of power yields a better test. Our test performs better in power comparison than the test proposed by Xiong et al. (2021) \cite{xiongetal}.

	\begin{center}
		\noindent{\bf Table 8}.  \small{{Powers of  $\boldsymbol{\hat{\Delta}}_{2,2}$, $T_1$ and $T_2$ for $N=20,50, 100$ at significance level $\alpha=0.05$}}
		
		\resizebox{!}{!}{
			\begin{tabular}{ p{1.0cm} p{1.0cm} p{1.0cm} p{1.0cm} p{1.0cm} p{1.0cm} p{1.0cm} } 
				\hline
				\ &\ & $T_1$  & $T_2$ & $\boldsymbol{\hat{\Delta}}_{2,2}$  & $T_1$ & $\boldsymbol{\hat{\Delta}}_{2,2}$  \\
				\hline
				$N$ & m  & $\chi^2_{(2)}$ & $\chi^2_{(2)}$ & $\chi^2_{(2)}$  &  $N(0,1)$ & $N(0,1)$ \\
				\hline
				\vspace{0.001cm}
				\ & 2   & 0.3999  & 0.4765  & 0.8861 & 0.0549 & 0.0231 \\
				\ & 3   & 0.5133  & 0.6842  & 0.8976 & 0.0501 & 0.0235  \\
				20 & 4  & 0.5962 & 0.5082  & 0.8956 & 0.0501 & 0.0224  \\
				\ & 6  & 0.6157  & 0.5187  & 0.8641  & 0.0440 & 0.0244  \\
				\ & 7   & 0.6234  & 0.4431  & 0.8889 & 0.0535 & 0.0253  \\
				\ & 8   & 0.5650  & 0.5673  & 0.8814 & 0.0518 & 0.0230  \\
				\hline
				\ & 5   & 0.9999  & 0.9813 & 0.9988 & 0.0569 & 0.0199 \\
				50 & 8   & 0.9995 & 0.9874  & 0.9990 & 0.0558 & 0.0233  \\
				\ & 20   & 0.9971 & 0.9770  & 0.9979 & 0.0494 & 0.0226 \\
				\hline
				\ & 8     & 0.9989  & 0.9936  & 1.0000 & 0.0497 & 0.0288  \\
				100 & 10   & 0.9992  & 0.9954  & 1.0000 & 0.0509 & 0.0307  \\
				\ & 15    & 0.9987  & 0.9994  & 1.0000 & 0.0541 & 0.0239  \\
				\ & 20     & 0.9978  & 0.9974  & 1.0000 & 0.0513 &  0.0231  \\
				
				\hline
		\end{tabular}}
	\end{center} \label{table8}

	We may therefore conclude that our suggested test, which is based on the cumulative past and residual extropy of the $n$th lower and upper $k$-record value, works satisfactorily in the simulation research. Our test perform better than Park (1999) \cite{park}, Xiong et al.(2021) \cite{xiongetal} and Jose and Sathar (2022) \cite{josesathar} in power comparison. We, therefore, anticipate that the proposed test will be superior to the competing tests in many real-world applications.
	
	\section{ Real data application }
	Jose and Sathar (2022) also used dataset 1 for their proposed test of symmetry. Dataset 1 from Montgomery et al. (2021) \cite{montigomary} has a normal distribution (symmetric model) as a suitable model.\\
	
	Dataset 1: 15.5, 23.75, 8.0, 17.0, 5.5, 19.0, 24.0, 2.5, 7.5, 11.0, 13.0, 3.75, 25.0,9.75, 22.0, 18.0, 6.0, 12.5, 2.0, 21.5.\\
	
	The normal distribution is symmetric. This fact is verified by our test. The value of the test statistics $\boldsymbol{\hat{\Delta}}_{2,2}$ is 0.1531 with an estimated $p$-value 0.2969 when window size $m=2$ and sample size $N=20.$ Our test based on $\boldsymbol{\hat{\Delta}}_{2,2}$ fail in rejecting null hypothesis even if the significance level is $10\%$. That is because the dataset has normal distribution as a suitable model.
	
	Xiong et al. (2022) \cite{xiongetal} used dataset 2 for their proposed test of symmetry. Dataset 2 from Qiu and Jia (2018b) \cite{quijia2018b} represent active repair times (in hours) for an airborne communication transceiver.  \\
	
	Dataset 2: 0.2, 0.3, 0.5, 0.5, 0.5, 0.5, 0.6, 0.6, 0.7, 0.7, 0.7, 0.8, 0.8, 1.0, 1.0, 1.0, 1.0,1.1, 1.3,1.5,1.5, 1.5, 1.5, 2.0, 2.0, 2.2, 2.5, 3.0, 3.0, 3.3, 3.3, 4.0, 4.0, 4.5, 4.7, 5.0, 5.4, 5.4, 7.0, 7.5, 8.8, 9.0, 10.3, 22.0, 24.5. \\
	
	This data can be fitted by inverse Gaussian (IG) distribution as pointed out in Qiu and Jia (2018b) \cite{quijia2018b}. IG distribution is not symmetric see, Xiong et al (2021) \cite{xiongetal} and Qiu and Jia (2018b) \cite{quijia2018b}. This fact is verified by our test. The value of the test statistics $\boldsymbol{\hat{\Delta}}_{2,2}$ is 3.6678 with an estimated p-value 0 when window size m=20 and sample size $N=45.$ Our test based on $\boldsymbol{\hat{\Delta}}_{2,2}$ succeed
	in rejecting the null hypothesis even if the significance level is small enough, say, $1\%$. This further shows the advantage of our test.\\
	
	Dataset 3: 1.42, 0.84, 2.32, 1.84, 2.4, 0.9, 1.49, 0.87, 1.36, 1.25, 1.25, 1.8, 0.86, 0.04, 0.49, 2.08, 0.58, 0.22, 0.06, 1.7, 2.67, 2.39, 2.32, 2.98, 3.21, 1.99, 1.3, 1.25, 1.76, 1.67, 1.36, 1.57, 1.21, 1.24, 1.62, 0.93, 1.32, 0.86, 1.48, 0.85, 1.23, 1.23, 2.14. \\
	
	Dataset 3 is taken from Sathar and Jose (2020) \cite{sathar2020b}. Sathar and Jose (2020) \cite{sathar2020b} proposed Normal distribution (symmetric model) as a suitable model for this data set. Jose and Sathar (2021) \cite{josesathar} also used this dataset in testing symmetry.\\
	
	Dataset 4: 99, 61, 86, 113, 96, 99, 83, 57, 80, 79, 75, 70, 15, 62, 87, 95, 81, 71, 44, 13, 52,	97, 146, 52, 52, 29, 108, 135, 102, 48, 66, 90, 22, 72, 176, 107, 84, 83, 37, 67, 83, 36, 49, 39, 102, 66, 154, 72, 63, 83, 77.\\
	
	Dataset 4 is taken from Thomas and Jose (2021) \cite{thomasjose21}. Thomas and Jose (2021) \cite{thomasjose21} proposed Burr-type XII distribution (skewed model)	as a suitable model for this data set. Jose and Sathar (2021) \cite{josesathar} also used this dataset in testing symmetry.\\
	
	Dataset 5 : 0.0518, 0.0518, 0.1009, 0.1009, 0.1917, 0.1917, 0.1917, 0.2336, 0.2336, 0.2336, 0.2733, 0.2733, 0.3467, 0.3805, 0.3805, 0.4126, 0.4431, 0.4719, 0.4719, 0.4993, 0.6162, 0.6550, 0.6550, 0.7059, 0.7211, 0.7356, 0.7623, 0.7863, 0.8178, 0.8810, 0.9337, 0.9404, 0.9732, 0.9858.\\
	
	Dataset 5 is Transformed vinyl chloride data into  uniform
	distribution using  probability integral transformation see, Xiong et al (2022) \cite{xiongetal20}. \\
	
	Dataset 6 : 0.014, 0.034, 0.059, 0.061, 0.069, 0.080, 0.123, 0.142, 0.165, 0.210, 0.381, 0.464, 0.479, 0.556, 0.574, 0.839, 0.917, 0.969, 0.991, 1.064, 1.088, 1.091, 1.174, 1.270, 1.275, 1.355, 1.397, 1.477, 1.578, 1.649, 1.702, 1.893, 1.932, 2.001, 2.161, 2.292, 2.326, 2.337, 2.628, 2.785, 2.811, 2.886, 2.993, 3.122, 3.248, 3.715, 3.790, 3.857, 3.912, 4.100.\\
	
	We considered dataset 6 from Lawless (2011) \cite{law11} which represents the quantity of 1000 cycles to failure for electrical appliances in a life test. 
	
	See Table  \ref{table8} for the value of test statistics and p-value for different datasets based on the specific window size and sample size of each dataset. Jose and Sathar (2021) \cite{josesathar} also used this dataset in testing symmetry.\\

	\vspace{0.4cm}
	\begin{center}
		\noindent{\bf Table 9}.\small{{ Description of models fitted }} 
		
		\resizebox{!}{!}{
			\begin{tabular}{ p{2.0cm} p{1.0cm} p{1.0cm} p{1.0cm} p{2.0cm}  } 
				\hline
				Dataset & N  & m & $\boldsymbol{\hat{\Delta}}_{2,2}$ & p-value  \\
				\hline 
				
				Dataset 1  & 20  & 2  & 0.1531 & 0.2969 \\
				Dataset 2  & 45  & 20 & 3.6678 & 0.0000 \\	
				Dataset 3  & 43  & 3  & 0.1545 & 0.2821 \\
				Dataset 4  & 51  & 25 & 6.2144 & 0.0000 \\
				Dataset 5  & 34  & 11 & 0.0247 & 0.4425  \\
				Dataset 6  & 50  & 2  & 0.5776 & 0.0210 \\
				\hline
		\end{tabular}}
	\end{center}\label{table9}
	If we are testing at a 5\% level of significance then the $p$-value less than 0.0500 detects asymmetric nature and a $p$-value more than 0.005 detects the symmetric nature of data.  Table 9 confirms that the newly proposed test determines whether the random sample's distribution is symmetric or asymmetric. The p-values indicate that, at a 5\% level of significance, Datasets 2, 4, and 6 do not have symmetry in the distribution of the random sample. Similar to this, a moderate p-value suggests accepting symmetry in the distribution of Datasets 1, 3, and 5. As a result, we could verify that the test statistic correctly identified the symmetry in the random variable's distribution.
	
	\section{Conclusion and future work}
	Gupta and Chaudhary (2022) \cite{guptaskc} proved cumulative past extropy of $n$th lower $k$-record value is equal to cumulative residual extropy of $n$th upper $k$-record value if and only if the underlying distribution is continuous symmetric distribution.	Using the above result, we proposed a new test for symmetry. We provided critical value and power against $\chi^{(1)}$ distribution for different sample size at significance level $\alpha=0.10$, $\alpha=0.05$ and $\alpha=0.01$. We calculated power of test
	against $\chi^{(1)}$ ,$\chi^{(2)}$, $\chi^{(3)}$ and $N(0,1)$ at 5\% significance level for sample size 20,50 and 100. Power comparison is done with competitors and our test performs better. We applied our test over six real-life examples and proposed a test that detects symmetric or asymmetric nature well with a significant p-value.
	
	One may use some other characteristics provided by Gupta and Chaudhary (2022), of symmetric distribution in testing symmetry using extropy. Also, one may do testing of exponentiality, uniformity and normality based on characteristics of exponential, uniform and normal distribution using cumulative past and residual extropy of record values.\\
	\\
	\\
	\noindent \textbf{ \Large Funding} \\
	\\
	Santosh Kumar Chaudhary would like to thank the Council Of Scientific And Industrial Research (CSIR), Government of India ( File Number 09/0081(14002)/2022-
	EMR-I ) for financial assistance.\\
	\\
	\textbf{ \Large Conflict of interest} \\
	\\
	The authors declare no conflict of interest.\\

	\section*{Appendix}
	1.	The following steps were used to determine the critical values and compute the power of our proposed test and that of other tests for symmetry at significance level $\alpha=0.10, \ \alpha=0.05, \alpha=0.01:$\\
	(1) we defined a function to calculate absolute value of $\boldsymbol{\hat{\Delta}}_{2,2}$ .\\
	(2) Generate a sample of size $N$ from the standard normal distribution and compute the test statistics for the sample data;\\
	(3) Repeat Step 1 for 10,000 times and determine the 950th, 975th and 995th quantile respectively of the test statistics as the critical value;\\
	(4) Generate a sample of size $N$ from the null distribution and check if the absolute value of the test statistic is greater than the critical value;\\
	(5) Repeat Step 3 for 10,000 times and the percentage of rejection is the power of the test.

	\noindent 2. Python code for critical value and power of the test statistics against a chi-square distribution with 1 degree of freedom. The value of m and N can be changed according to the need of study.\\
	
	import numpy as np
	
	def calD2(sample, m, N):
	
	\ \ \ \ \ sample.sort()
	
	\ \ \ \ \ Junx=-1.0/2/N*sum([(1-2*np.log(1-i/(N+1)))**2*(1-i/(N+1))**4* (sample[min(i+m-1,N-1)] - sample[max(i-m-1,0)]) for i in range(1,N+1)]) /(2*m/N)
	
	\ \ \ \ \ Jlnx=-1.0/2/N*sum([(1-2*np.log(i/(N+1)))**2*(i/(N+1))**4*(sample [min(i+m-1,N-1)]- sample[max(i-m-1,0)]) for i in range(1,N+1)])/(2*m/N)
	
	\ \ \ \ \ D2=Junx-Jlnx
	
	\ \ \ \ \ return abs(D2)
	
	list1=[]
	
	m=2
	
	N=10
	
	for i in range(10000):
	
	\ \ \ \ \ sample1=np.random.normal(0.0,1.0,N)
	
	\ \ \ \ \ list1.append(calD2(sample1, m, N))
	
	print(list1)
	
	print("critical value for alpha=0.10 is ",np.quantile(list1, 0.950))
	
	print("critical value for alpha=0.05 is ",np.quantile(list1, 0.975))
	
	print("critical value for alpha=0.01 is ",np.quantile(list1, 0.995))

	count2=0
	
	for i in range(10000):
	
	\ \ \ \ \ sample2=np.random.chisquare(1,N)
	
	\ \ \ \ \ if calD2(sample2, m, N)>np.quantile(list1, 0.950):
	
	\ \ \ \ \ \ \ \ \ \ count2=count2+1
	
	print("power when alpha=0.10",count2/10000)

	count3=0
	
	for i in range(10000):
	
	\ \ \ \ \ sample3=np.random.chisquare(1,N)
	
	\ \ \ \ \ if calD2(sample3, m, N)>np.quantile(list1, 0.975):
	
	\ \ \ \ \ \ \ \ \ \ count3=count3+1
	
	print("power when alpha=0.05",count3/10000)

	count4=0
	
	for i in range(10000):
	
	\ \ \ \ \ sample4=np.random.chisquare(1,N)
	
	\ \ \ \ \ if calD2(sample4, m, N)>np.quantile(list1, 0.995):
	
	\ \ \ \ \ \ \ \ \ \ count4=count4+1
	
	print("power when alpha=0.01",count4/10000) \\
	
	\noindent 3. Python code to calculate the value of test statistics and p-value corresponding to dataset 1 as an alternative. Sample 2 is real life dataset as an alternative.
	
	import numpy as np
	
	def calD2(sample, m, N):
	
	\ \ \ \ \ sample.sort()
	
	\ \ \ \ \ Junx=-1.0/2/N*sum([(1-2*np.log(1-i/(N+1)))**2*(1-i/(N+1))**4* (sample[min(i+m-1,N-1)] - sample[max(i-m-1,0)]) for i in range(1,N+1)]) /(2*m/N)
	
	\ \ \ \ \ Jlnx=-1.0/2/N*sum([(1-2*np.log(i/(N+1)))**2*(i/(N+1))**4*(sample [min(i+m-1,N-1)]- sample[max(i-m-1,0)]) for i in range(1,N+1)])/(2*m/N)
	
	D2=Junx-Jlnx
	
	return D2
	
	m=2
	
	N=20
	
	count2=0
	
	for i in range(10000):
	
	sample1=np.random.normal(0.0,1.0,N)
	
	sample2=[15.5, 23.75, 8.0, 17.0, 5.5, 19.0, 24.0, 2.5, 7.5, 11.0, 13.0, 3.75, 25.0, 9.75, 22.0, 18.0, 6.0, 12.5, 2.0, 21.5]
	
	if calD2(sample2, m, N) < calD2(sample1, m, N):
	
	count2=count2+1
	
	print(count2/10000) \\
	
	\noindent 4. Python code for empirical density of test statistics
	
	import pandas as pd
	
	from matplotlib import pyplot
	
	from numpy.random import normal
	
	from numpy import hstack
	
	import matplotlib.pyplot as plt
	
	import seaborn as sns
	
	import numpy as np
	
	def calD2(sample, m, N):
	
	\ \ \ \ \ sample.sort()
	
	\ \ \ \ \ Junx=-1.0/2/N*sum([(1-2*np.log(1-i/(N+1)))**2*(1-i/(N+1))**4* (sample[min(i+m-1,N-1)] - sample[max(i-m-1,0)]) for i in range(1,N+1)]) /(2*m/N)
	
	\ \ \ \ \ Jlnx=-1.0/2/N*sum([(1-2*np.log(i/(N+1)))**2*(i/(N+1))**4*(sample [min(i+m-1,N-1)]- sample[max(i-m-1,0)]) for i in range(1,N+1)])/(2*m/N)
	
	D2=Junx-Jlnx
	
	return D2
	
	list1=[]
	
	m=45
	
	N=100
	
	for i in range(10000):
	
	sample1=np.random.normal(0.0,1.0,N)
	
	list1.append(calD2(sample1, m, N))
	
	print(list1)
	
	sns.distplot(list1, hist=False, kde=True, bins=int(250/5), color = 'darkblue', hist\_kws={'edgecolor':'black'}, kde\_kws={'linewidth': 3})\\

	\vspace{5cm}
	
	\noindent
	{\bf Santosh Kumar Chaudhary}\\
	Department of Mathematics,\\
	Indian Institute of Technology Kharagpur\\
	Kharagpur-721302, INDIA\\
	E-mail: skchaudhary1994@kgpian.iitkgp.ac.in \\

	\noindent
	{\bf  Nitin Gupta} \\
	Department of Mathematics,\\
	Indian Institute of Technology Kharagpur\\
	Kharagpur-721302, INDIA\\
	E-mail: nitin.gupta@maths.iitkgp.ac.in\\


\begin{thebibliography}{99}
		
		
		\bibitem[1]{xiongetal} Xiong, P., Zhuang, W., and Qiu, G. (2021), Testing symmetry based on the extropy of record values, Journal of Nonparametric Statistics, 33:1, 134-155.
		\bibitem[2]{xiongetal20} Xiong, P., Zhuang, W., and Qiu, G. (2022), Testing exponentiality based on the extropy of record values. J Appl Stat.  Oct 31;49(4):782-802.			
		\bibitem[3]{vasicek} Vasicek, O. ( 1976), A test for normality based on sample entropy. J R Stat Soc Ser B Methodol.38:54–59. 
		\bibitem[4]{park} Park, S. (1999), ‘A Goodness-of-fit Test for Normality Based on the Sample Entropy of Order Statistics’, Statistics and Probability Letters, 44, 359–363.
		\bibitem[5]{guptaskc} Gupta, N., and Chaudhary, S. K. (2022), Some characterizations of continuous symmetric distributions are based on the extropy of record values. arXiv preprint arXiv:2208.07116 (Communicated with Journal).
		\bibitem[6]{josesathar} Jose, J., and Sathar, E.I.A. (2022), Symmetry being tested through simultaneous application of upper and lower $k$-records in extropy, Journal of Statistical Computation and Simulation, 92:4, 830-846.
		\bibitem[7]{ahmadi}	Ahmadi, J. (2021), Characterization of continuous symmetric distributions using information measures of records. Stat Papers 62, 2603–2626.
		\bibitem[8]{fashandiahmadi12} Fashandi, M., and Ahmadi, J. (2012), Characterizations of symmetric distributions based on Rényi entropy. Stat Probab Lett 82:798–804.
		\bibitem[9]{mcwill}	McWilliams T.P. (1990), A distribution-free test for symmetry based on a run statistic. J Am Stat Assoc. ;85(412):1130–1133. 
		\bibitem[10]{chengbala04} Cheng, W.H., and Balakrishnan, N. (2004), A modified sign test for symmetry. Commun Stat Simul Comput. 33(3):703–709.
		\bibitem[11]{corzobaba} Corzo, J., and Babativa, G. (2013), A modified runs test for symmetry. J Stat Comput Simul. 83(5):984–991.
		\bibitem[12]{modagast} Modarres, R., and  Gastwirth, J.L. (1996), A modified runs test for symmetry. Stat Probab Lett. ;31(2):107–112.
		\bibitem[13]{lad15} Lad, F., Sanfilippo,  G., and Agro. G. (2015), Extropy: Complementary dual	of entropy. Statistical Science 30 (1):40–58.
		\bibitem[14]{montigomary} Montgomery, D.C., Peck, E.A., and  Vining, G.G. (2021), Introduction to linear regression analysis. New York: John Wiley and Sons.
		\bibitem[15]{quijia2018b} Qiu, G., and Jia, K. (2018b), ‘Extropy Estimators with Applications in Testing Uniformity’, Journal of Nonparametric Statistics, 30, 182–196.
		\bibitem[16]{shanon48} Shannon, C. E. (1948), A mathematical theory of communication, Bell System Tech. J.  27: 379-423, 623-656.
		\bibitem[17]{jittojoseexp22} Jose, J., and Sathar, E.I.A. (2022), Characterization of exponential distribution using extropy based on lower k-records and its application in testing exponentiality,
		Journal of Computational and Applied Mathematics, 402, 113816, 0377-0427.
		\bibitem[18]{sathar2020b}  Sathar E.I.A., and  Jose J. (2020b), Past extropy of $k$-records. Econ Qual Control. 35(1):25–38.
		\bibitem[19]{thomasjose21} Thomas, P.Y., Jose, J. (2021), On Weibull–Burr impounded bivariate distribution. Jpn J Stat Data Sci.4(1):73–105.
		\bibitem[20]{josethomus18} Jose J., and Thomas P.Y. (2018), A new bivariate distribution with extreme value type I and Burr type XII distributions as marginals. J Kerala Stat Assoc. 29:1–24.
		\bibitem[21]{law11} Lawless, and J.F. (2011), Statistical Models and Methods for Lifetime Data, vol. 362, Wiley, Hoboken.
		\bibitem[22]{dzko1976} Dziubdziela W., and Kopocinski B. (1976), Limiting properties of the $k$th record values,  Applications Mathematicae 2 (15) 187–190.
		\bibitem[23]{ahsan95} Ahsanullah, M. (1995), Record Statistics. Nova Science Publishers, New York.
		\bibitem[24]{arnoldbalanag98} Arnold, B. C., Balakrishnan, N., and Nagaraja. H. N. (1998), Records, vol. 768. New York: John Wiley and Sons. 
		\bibitem[25]{arnoldbalanaga08} Arnold C., Balakrishnan N., and Nagaraja H. N. (2008). A first course in order statistics, SIAM .
		\bibitem[26]{ahsan2004} Ahsanullah, M., (2004), Record values–theory and applications, University Press of America.
		\bibitem[27]{gibchakr92} Gibbons, D., and Chakraborti, S. (1992), Nonparametric Statistical Inference, New York: Marcel Dekker.
		\bibitem[28]{taj94} Tajuddin, I. (1994), ‘Distribution-free Test for Symmetry Based on Wilcoxon Two-sample Test’, Journal of Applied Statistics, 21(5), 409–415.
		\bibitem[29]{baklizi03} Baklizi, A. (2003), ‘A Conditional Distribution Free Runs Test for Symmetry’, Journal of Nonparametric Statistics, 15(6), 713–718.
		\bibitem[30]{baklizi07} Baklizi, A. (2007), ‘Testing Symmetry Using a Trimmed Longest Run Statistic’, Australian and New Zealand Journal of Statistics, 49(4), 339–347. 
		\bibitem[31]{baklizi08} Baklizi, A. (2008), ‘Improving the Power of the Hybrid Test’, International Journal of Contemporary Mathematical Sciences, 3(10), 497–499.
	\end{thebibliography}
\end{document}